\documentclass[10pt,twocolumn,twoside]{article}

\usepackage[letterpaper,top=0.75in,bottom=0.75in,left=1in,right=1in,columnsep=13.5pt,footskip=0.3in]{geometry}
\usepackage[english]{babel}
\usepackage{lmodern}
\usepackage[scaled]{helvet}
\usepackage[T1]{fontenc}
\usepackage{times}
\usepackage{amsmath,amssymb}
\usepackage{bbold}
\usepackage{graphicx}
\usepackage{booktabs}
\usepackage{xcolor}
\usepackage{mdframed}
\usepackage{float}
\usepackage[explicit]{titlesec}
\usepackage{fancyhdr}
\usepackage[labelfont={bf,sf},labelsep=period,figurename=Fig.]{caption}
\usepackage[numbers,sort&compress,merge,round]{natbib}
\usepackage[nopatch=eqnum]{microtype}
\usepackage[colorlinks=true,allcolors=blue]{hyperref}

\newcommand{\hexagon}{\mbox{\fontencoding{U}\fontfamily{wasy}\fontseries{m}\selectfont\char55}}

\definecolor{sigtext}{RGB}{0,101,165}
\definecolor{sigback}{RGB}{205,217,235}
\newcommand{\titlefont}{\fontfamily{lmss}\bfseries\fontsize{22pt}{24pt}\selectfont}
\newcommand{\authorfont}{\normalfont\sffamily\bfseries\fontsize{10}{12}\selectfont}
\newcommand{\affilfont}{\normalfont\sffamily\fontsize{8}{9.5}\selectfont}
\newcommand{\absfont}{\normalfont\sffamily\bfseries\fontsize{9}{12}\selectfont}
\newcommand{\keywordsfont}{\normalfont\rmfamily\fontsize{8}{10}\selectfont}
\newcommand{\footerfont}{\normalfont\sffamily\fontsize{8}{10}\selectfont}

\titleformat{name=\section,numberless}{\large\sffamily\bfseries}{}{0em}{#1}
\titleformat{name=\subsection,numberless}[runin]{\sffamily\bfseries}{}{0em}{#1. }
\titlespacing*{\section}{0pc}{3ex plus4pt minus3pt}{5pt}
\titlespacing*{\subsection}{0pc}{2.5ex plus3pt minus2pt}{2pt}

\newenvironment{materialsandmethods}
  {\par\vskip 3.25ex plus 1ex minus .2ex
   \noindent{\sffamily\normalsize\bfseries Materials and Methods}\par\nobreak
   \vskip 1.5ex plus .2ex\small}
  {\par}
\newenvironment{acknowledgments}
  {\par\vskip 3.25ex plus 1ex minus .2ex
   \noindent{\sffamily\normalsize\bfseries ACKNOWLEDGMENTS.\hspace{1.5ex plus .2ex}}\small\ignorespaces}
  {\par}

\renewcommand{\bottomfraction}{0.8}

\renewcommand{\bibfont}{\normalfont\sffamily\fontsize{7}{9}\selectfont}
\renewcommand{\bibnumfmt}[1]{#1.}

\makeatletter
\let\amseqref\eqref
\renewcommand{\eqref}[1]{\textup{{\normalfont Eq.~(\ref{#1}}\normalfont)}}
\renewcommand\tagform@[1]{\maketag@@@{[\ignorespaces#1\unskip\@@italiccorr]}}
\makeatother

\fancypagestyle{firstpage}{%
  \fancyhf{}%
  \fancyfoot[R]{\footerfont Preprint\hspace{7pt}|\hspace{7pt}\textbf{\thepage\textendash\pageref{mainlastpage}}}}
\fancypagestyle{supplement}{\fancyhf{}\fancyhead[R]{\thepage}}

\newfloat{sigstatement}{b!}{sst}

\hypersetup{pdftitle={A Tissue Vertex Model grounded in Continuum Mechanics},pdfauthor={Jairo M. Rojas, Mayisha Z. Nakib, Sascha Hilgenfeldt}}

\begin{document}

\twocolumn[{%
  {\raggedright\titlefont A Tissue Vertex Model grounded in Continuum Mechanics\par}
  \vskip10pt
  {\raggedright\authorfont Jairo M. Rojas\textsuperscript{a,b}, Mayisha Z. Nakib\textsuperscript{a,b}, and Sascha Hilgenfeldt\textsuperscript{a,b,1}\par}
  \vskip4pt
  {\raggedright\affilfont \textsuperscript{a}Department of Physics, The Grainger College of Engineering, University of Illinois, Urbana-Champaign, Illinois 61801, USA; \textsuperscript{b}Mechanical Science and Engineering, The Grainger College of Engineering, University of Illinois, Urbana-Champaign, Illinois 61801, USA\par}
  \vskip12pt}]
\thispagestyle{firstpage}

{\noindent\absfont
Practical modeling of disordered single-layer confluent tissues has relied heavily on the two-dimensional standard vertex model (SVM), which simplifies cell elasticity using an energy functional determined by quadratic dependence of cell area and perimeter values. Realistically, even a coarse-grained cell interior consists of effective continuum materials representing cytoplasm and peripheral cortex, with elastic energy depending on stress and strain fields. It is shown that this energy is analytically tractable for polygonal cell shapes, that its area dependence can be eliminated, that it correlates well with cell perimeter length, and that the perimeter dependence of the correlation is not quadratic. This results in a revised vertex model derived from continuum elasticity whose parameters are predicted from tissue material properties and whose resultant forces are far less strain-dependent than in the SVM.
\par}
\vskip10pt
{\noindent\keywordsfont Continuum Mechanics of Cells$|$ Vertex Models $|$ Soft Materials
$|$ Tissue Mechanics\par}
\vskip12pt

\begin{figure}[!b]
    \centering
\includegraphics[width=0.38\textwidth]{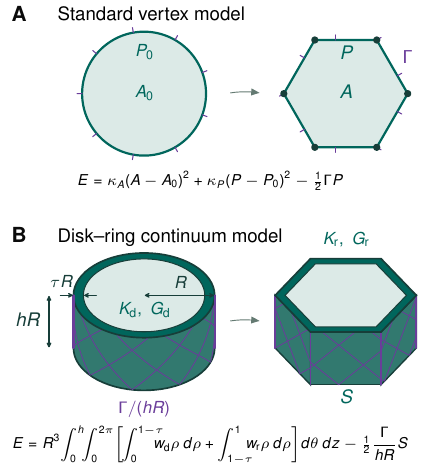}
    \caption{(A) The standard vertex model (SVM) relates elastic deformation energy in a polygonal cell to its area and perimeter relative to a (circular) reference state, using ad-hoc quadratic dependences. (B) The  model developed here instead computes energy from elastic continuum mechanics via complete stress and strain fields of a polygonal cell in polar coordinates ($\rho=r/R,\theta$). The unstrained reference state is a circular cylinder (cytoplasm) surrounded by an annulus (cortex), each with their own  material properties. Both models assume uniform adhesion energies per lateral length $\Gamma$, contributing proportional to perimeter $P$ and mantle area $S$, respectively.}
    \label{fig:svm}
\end{figure}
\begin{sigstatement}[b]
  \begin{mdframed}[linewidth=0.7pt,backgroundcolor=sigback,linecolor=sigtext,fontcolor=sigtext,innertopmargin=6pt,innerrightmargin=6pt,innerbottommargin=6pt,innerleftmargin=6pt,skipabove=0pt,skipbelow=0pt]
    {\sffamily\large\bfseries Significance Statement\par}
    \vskip5pt
    {\sffamily\small
Modeling of physical and mechanical properties of confluent biological tissues, such as epithelia, has relied heavily on the concept of {\em vertex models}, in which cell shapes are simplified as polygons and their elastic mechanical energies are assumed as ad-hoc quadratic functions of area and perimeter. Questioning these ubiquitous assumptions, we derive elastic energies instead from continuum mechanics and measured material properties of cell cytoplasm and cell cortex. It is found that these energies do correlate well with cell perimeter changes, putting the concept of vertex models on a firm footing in continuum theory for the first time. However, the perimeter correlation is not quadratic, but instead close to linear, while area variations play a negligible role. These findings are robust for a wide range of realistic material properties and alter modeling predictions for the distributions of intercellular forces, and consequently, for the morphology of healthy or diseased tissue. Revised vertex models with these modifications are no more expensive computationally than the standard approach, while being fully compatible with continuum mechanics of soft matter.
    \par}
  \end{mdframed}
  \vskip3pt
  {\sffamily\scriptsize
  J.M.R. and S.H. designed the theoretical research; J.M.R. and M.Z.N. performed theoretical research and analyzed data; J.M.R. and S.H. wrote the paper.\par\vskip5pt
  The authors declare no competing interest.\par\vskip5pt
  \textsuperscript{1}To whom correspondence should be addressed. E-mail: sascha@illinois.edu\par}
\end{sigstatement}

Despite deep insights into biological processes in living cells based on genetics, protein networks, and biochemical pathways, our understanding of the {\em mechanical} behavior of cells and tissues as soft materials is lacking. Whether tissues are rigid or floppy, elastic or viscoelastic, speaks to the biological function of healthy tissues as well as to pathological changes in diseased organs. It is very difficult to realistically describe the mechanics of such enormously complex materials. Simpler modeling descriptions hope to capture the gist of the physics and mechanics, allowing for the identification of morphological indicators that translate into mechanical properties and, thus, to tissue health diagnostics \cite{Bi2015,park2015unjamming,Kim18,Kim19,yan2019multicellular}. Finding a balance between intricate models impractical for drawing strong general conclusions and oversimplified models of questionable connection with real systems has been a challenge.

\subsection*{The Standard Vertex Model}
For a number of years, a radically simplifying approach to the modeling of elastic mechanics in confluent tissue layers has been widely used, in which the elastic energy of a cell (modeled as a two-dimensional (2D) polygon) is a function of its area $A$ and perimeter $P$ only (projected into the plane of the tissue). These {\em vertex models} allow for simple simulations with many cells \cite{Farhadifar2007,Staple10,Kim15,Kim19,kim2021embryonic,bandil2023continuum,alt2017vertex}. The mechanical energy functional has been almost invariably chosen as the "Standard Model" introduced in early pioneering work \cite{Farhadifar2007,Staple10}, where the elastic energy $U$ of a cell is (cf.\ Fig.~\ref{fig:svm}A)
\begin{equation}
\begin{aligned}
U & = &  \frac{1}{2}\kappa_A (A-A_0)^2 + \frac{1}{2}\kappa_P (P-P_0)^2 \\
& = &  \frac{1}{2}\kappa_A (A-A_0)^2 + \frac{1}{2}\kappa_P P_0^2 {\mathcal P}^2\,.
\end{aligned}
\label{eq:svm}
\end{equation}
Here, the cell has unstrained equilibrium areas $A_0$ and perimeter $P_0$ -- these are typically taken as those of a circular disk, i.e.\ $P_0=(4\pi A_0)^{1/2}$, backed by findings from biology and bioengineering that unattached cells tend to round up to minimize their surface or (in 2D) their perimeter under cytoskeletal tension, in isolation \cite{coughlin1998tensegrity,clark2011mechanics} as well as within epithelia when detaching from neighbors \cite{hoijman2015mitotic}. Simulations with this $U$ together with a uniform adhesion energy between cells constitutes the Standard Vertex Model (SVM), in which researchers found intriguing loss-of-rigidity transitions that can be detected from average perimeter lengths \cite{Bi2015,Kim15}, while the area elasticity term in \eqref{eq:svm} is unimportant. For later use, we have defined the dimensionless perimeter excess ${\mathcal P}= (P-P_0)/P_0$. There are shortcomings of the model: the material parameters (moduli) $\kappa_A$ and $\kappa_P$ have no direct measurable correlate. More fundamentally, while the terms in \eqref{eq:svm} are "plausible" elastic terms in the sense of harmonic potentials (powers of 2) in the geometric variables, there is no a priori reason to assume that mechanical energy is related to $A$ and $P$ in this way. %

\subsection*{Continuum model of cytoplasm and cortex}
The mechanical details of the nucleus, organelles, cytoskeleton, and membranes in each cell are too complex for efficient modeling of a tissue with many cells. Even direct numerical schemes, which can deal with arbitrary cell shapes, need to simplify the mechanics using continuum assumptions \cite{chen2000cell,brodland2007new,runser2024simucell3d}. We describe a new approach to epithelial cell modeling respecting continuum mechanics descriptions of both the cytoplasm and cortex, while preserving the simplicity of vertex models. Any such model must include adhesion between cells, but we focus on elastic energy $U$ only, as our model leaves the adhesion assumptions of the SVM unchanged (uniform adhesive strength $\Gamma$ on the lateral faces of the cell, Fig.~\ref{fig:svm}).

\begin{figure}[!t]
    \includegraphics[width=0.45\textwidth]{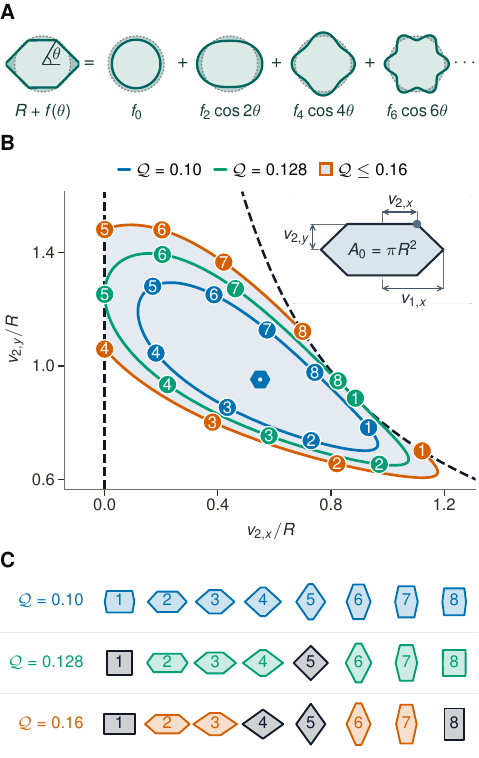}
    \caption{(A) Arbitrary deformations in the plane by  Fourier decomposition of the perimeter (outer edge of the deformed ring), each mode contributing $f_n \cos n\theta$ to the polar representation. (B) For given area, the shape of a symmetric hexagon is given by the vertex coordinates $v_{2,x}$ and $v_{2,y}$. We analyze all shapes in the shaded region, bounded by limits of convexity (dashed lines) and an upper limit for the shape measure ${\mathcal Q}={\mathcal Q}_{max}$ (red solid line). Other solid lines illustrate constant ${\mathcal Q}$, along which the shapes indicated in (C) are encountered.}
    \label{fig:shapes}
\end{figure}
\begin{table}[t]
 \caption{Biological reference values on cell geometry and material elasticity.
 See also Materials and Methods.
 }
 \label{tab:biological-baseline}
 \small
 \setlength{\tabcolsep}{3pt}
 \begin{tabular*}{\columnwidth}{@{\extracolsep{\fill}}lcrl@{}}
  \toprule
  Quantity & Symbol & Value & Sources \\
  \midrule
  cell radius & $R$ & $7.5\,\mu\mathrm{m}$ & \cite{Kaliman2016} \\
  cell height & $hR$ & $7\,\mu\mathrm{m}$ & \cite{Zehnder2015} \\
  cortical thickness & $\tau R$ & $0.30\,\mu\mathrm{m}$ & \cite{Janshoff2021Basal} \\
  \addlinespace
  cytoplasm bulk modulus & $K_{\mathrm d}$ & $3.2\,\mathrm{MPa}$ & \cite{Guo2017} \\
  cytoplasm shear modulus & $G_{\mathrm d}$ & $2.4\,\mathrm{Pa}$ & \cite{Guo2017} \\
  \addlinespace
  cortex bulk modulus & $K_{\mathrm r}$ & $0.53\,\mathrm{MPa}$ & \cite{Guo2017,cordes2020} \\
  cortex shear modulus & $G_{\mathrm r}$ & $5.1\,\mathrm{kPa}$ & \cite{Guo2017} \\
  \bottomrule
 \end{tabular*}
\end{table}
To arrive at a closed-form theory, we neglect shape variations along the apical-basal axis (though these can in principle be taken into account \cite{jairo2025}) and describe a cell as a cylindrical 3D elastic continuum of given height (Fig.~\ref{fig:svm}B). We acknowledge the different mechanical properties of cytoplasm (modeled as a circular cylinder of radius $R(1-\tau)$ and height $hR$ when unstrained) and cellular cortex (an annulus of thickness $\tau R$ when unstrained). The properties of both materials are described by elastic moduli; because of the uniformity along the cylinder axis, we explicitly model a 2D cross section as a disk (cytoplasm) surrounded by a ring (cortex). Epithelial tissues in experiment show little change in height with volume changes \cite{Zehnder2015}, so that we employ plane strain here. Assuming linear elasticity for clarity and analytical tractability (see below and the Supporting Information (SI) for a discussion of nonlinear effects), the Michell equations \cite{Michell1899,hulbert1963numerical,meleshko2003selected} provide solutions for deformations in polar coordinates $(\rho,\theta)$ when supplemented with boundary conditions of (i) an imposed shape $f(\theta)$ at the outer edge of the ring, (ii) continuity of displacements and stresses between disk and ring, and (iii) absence of tangential stresses at the outer edge of the ring (consistent with typical vertex model assumptions).  Strain $\epsilon_{ik}$ and stress $\sigma_{ik}$ fields ($i,k\in \{\rho,\theta\}$) follow directly, given shear and bulk moduli $G_d, K_d$ and $G_r, K_r$ for disk and ring, respectively. The elastic energy of the deformed cell is then analytically determined by integration of the local energy densities $w=\frac{1}{2}\sum_{i,k}\sigma_{ik}\epsilon_{ik}$ \cite{landau2020theory} over the disk and ring areas, i.e.,
\begin{equation}
U = R^3h \int_0^{2\pi} \left[\int_0^{1-\tau} w_d \rho d\rho + \int_{1-\tau}^{1} w_r \rho d\rho \right] d\theta\,,
    \label{eq:ucont}
\end{equation}
the 2D version of the elastic terms in Fig.~\ref{fig:svm}B. The linearity of the system allows decomposition into Fourier modes $n$ and the separate computation of energy contributions $U_n$ for each mode amplitude $f_n$, see Methods and the SI for details of these computations.

\begin{figure}[!b]
 \centering
 \includegraphics[width=\columnwidth]{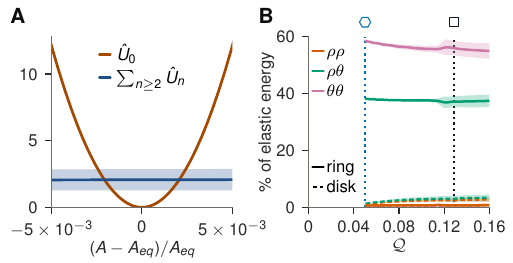}
 \caption{\label{fig:size-dependent-partition}Energy partition across the
 hexagonal shapes of Fig.~\ref{fig:shapes}. (A)~Relative energy contributions of mode $n=0$ (red) and all modes $n>0$ (blue) as area is varied relative to the area $A_{eq}$ minimizing total energy. Solid line is the mean, the shaded envelope marks standard deviations across the  shape family (these are smaller than the line thickness for $n=0$). (B)~Relative contributions to elastic energy (at equilibrium area) as a function of shape excess ${\mathcal Q}$. Cortex (ring) contributions are solid and dominate the dashed cytoplasm (disk) contributions. Cortical hoop ($\theta\theta$)  and shear ($\rho\theta$) contributions are largest and comparable. Vertical dashed lines mark the shape excess of the regular hexagon and the square (as degenerate hexagon).
 }
 \label{fig:modeimportance}
\end{figure}
Global quantities like area and perimeter of the cell can also be directly computed and their relationship with $U$ scrutinized. We do so for a wide variety of shape deformations of the unstrained circle into straight-edged polygons (representing the outer edge of the ring), parametrized geometrically. To scan a complete shape space, our approach preserves symmetry of the polygons along the $x$- and $y$-axes, restricting the non-zero Fourier modes to even $n$, cf.\ Fig.~\ref{fig:shapes}A.
We explain the parametrization of hexagons in detail here and demonstrate below that taking into account polygons with different edge number or without symmetry leaves the results nearly unchanged.
Symmetric straight-edged hexagons are exhaustively parametrized by their size (area $A$) and the coordinates of one vertex, see Fig.~\ref{fig:shapes}B. For fixed $A=A_0$, Fig.~\ref{fig:shapes}B shows the bounds of the parameter space used, requiring convexity of the polygon and setting an upper threshold to perimeter elongation. As our approach allows for coupled area and perimeter changes, we here introduce the shape excess parameter $\mathcal Q\equiv (P-P_c)/P_c$, where $P_c=(4\pi A)^{1/2}$ is the circular perimeter at the current area $A$. Thus, ${\mathcal Q}$ remains a measure linear in perimeter, but encodes shape only, decoupled from size, as opposed to ${\mathcal P}$. In practice, the differences between ${\mathcal Q}$ and ${\mathcal P}$ are slight, see SI.
We constrain the shape excess to ${\mathcal Q}\leq {\mathcal Q}_{max}$, with ${\mathcal Q}_{max}=0.16$, a value motivated by the validity limits of linear elasticity (beyond ${\mathcal Q}_{max}$, the deformation gradients do not stay small everywhere), which also captures a majority of experimentally observed cell shapes in MDCK epithelia \cite{bill_data}. The regular hexagon has $\mathcal{Q}_{\hexagon}\approx 0.05$.

The material moduli of disk and ring are chosen according to measurements in the literature probing cytoplasm and cortex of epithelial and epithelial-derived cells on length and time scales appropriate for static elasticity \cite{Zehnder2015,Luchtefeld2020,Guo2017,cordes2020}, see Table~\ref{tab:biological-baseline}, with values taken directly from MDCK-II cells wherever possible (see Materials and Methods). As soft materials, cytoplasm and cortex are nearly incompressible ($G_j\ll K_j$ for $j=d,r$), suggesting that modes $n=0$ (which explicitly changes area) and $n>0$ (which preserve area to leading order) should be analyzed separately. Note that $G_d\not = 0$ (we do not assume the cytoplasm is a liquid), whereas $K_d$, while large, is not a typical bulk modulus of homogeneous condensed matter ($\gtrsim 10^9\,$Pa) because volume changes in tissue cells involve osmotic poroelasticity \cite{spring1982epithelial,Zehnder2015,Guo2017}.

\begin{figure}[!t]
 \centering
 \includegraphics[width=\columnwidth]{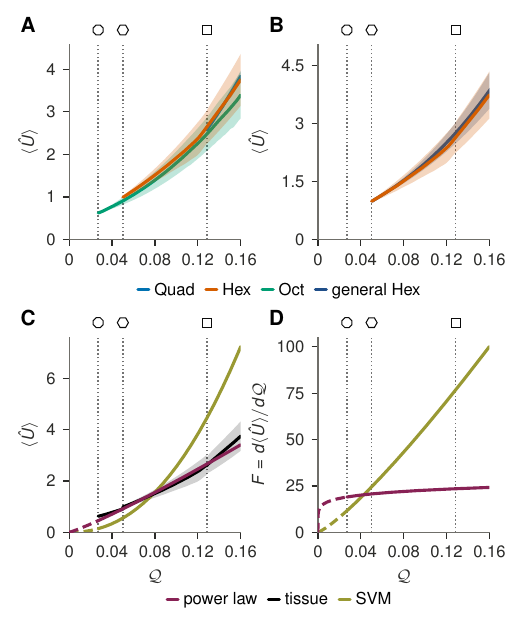}
 \caption{\label{fig:cross-family-summary}
 (A) Average energy-perimeter correlations (solid lines) for symmetric hexagonal (red), octagonal (green), and quadrilateral (blue) polygons, with shading indicating standard deviations. All topologies are consistently showing a slightly superlinear behavior in ${\mathcal Q}$. (B) The energy-perimeter correlation is not biased by the polygon symmetry, as a Monte-Carlo sampling of arbitrary hexagons (dark blue) agrees closely with the symmetric result (red). (C)
 Population-averaged energy (black), together with a pure power-law fit \eqref{UofPpower} (dark red) and the best fit (least square of relative errors) to the SVM quadratic model (yellow). (D) Derivatives of the model curves in (C), representing a scale of tensile force exerted on vertices in vertex model. The SVM forces are strongly perimeter dependent, while the continuum-based dependence is dramatically weaker.
 }
 \label{fig:anharmonic}
\end{figure}
Because of the near-incompressibility, for any given deformation total elastic energy is most sensitive to the $n=0$ mode. Leaving the two vertex coordinates of Fig.~\ref{fig:shapes} constant and varying $A$ identifies a sharp minimum of energy for a certain $A=A_{eq}$, which is shape-dependent, but always close to $A_0$. Figure~\ref{fig:modeimportance}A compares the contributions of $n=0$ and $n>0$ modes to elastic energy (normalized by the energy of a regular hexagon, $\hat{U}=U/U_{\hexagon}$) as $A$ varies through $A_{eq}$, demonstrating that at $A=A_{eq}$ the contribution of the $n=0$ mode is negligible. This rationalizes consistent findings of simulations using the SVM \cite{Kim15,Kim19}: area elasticity is unimportant when replaced by fixing the area at or near its equilibrium value.
Thus, it is sufficient to vary the vertex coordinates $v_x, v_y$ relative to $R$ (cf.\ Fig.~\ref{fig:shapes}B) to explore the space of deformations Fig.~\ref{fig:shapes}C.
In particular, we analyze the relationship between $U$ and the perimeter shape indicator ${\mathcal Q}$.
Figure~\ref{fig:modeimportance}B shows that, independent of ${\mathcal Q}$, the ring (cortex) contains most of the deformation energy, with the hoop ($\theta\theta$) and shear ($\rho\theta$) contributions of comparable size. All computations take into account modes up to $n_{\max}=78$, i.e., all half-wavelengths longer than the cortical ring thickness ($\pi/\tau\approx78.5$).

\subsection*{Anharmonic Perimeter Elasticity}
Figure~\ref{fig:anharmonic}A plots total elastic energies $\hat{U}$ vs ${\mathcal Q}$ using analytical computations across the entire space of hexagonal shapes (red curve). Rather than computing values for individual polygonal deformations, we
take into account all shapes with a fixed ${\mathcal Q}$ by integrating along a line of Fig.~\ref{fig:shapes}B, determining mean $\langle \hat{U}\rangle$ and standard deviation (assuming implicitly that all configurations have equal statistical weight).

The plot shows strong correlation of $\hat{U}$ and ${\mathcal Q}$ indicating that lengthening of perimeter does represent a reliable morphological indicator of increased elastic deformation energy, in agreement with the cortical concentration of energy in Fig.~\ref{fig:modeimportance}B.
Note that in a more realistic scenario less extreme shape deformations will be more probable, so that the standard deviations should be seen as upper limits, with the correlation likely even stronger for cell shapes practically encountered in confluent tissues.

\begin{figure}[!b]
 \centering
 \includegraphics[width=\columnwidth]{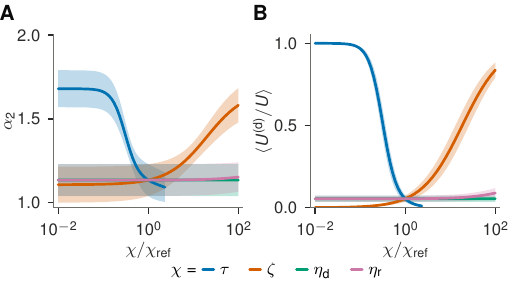}
 \caption{(A) Dependence of the best-fit exponent $\alpha_2$ in the elastic energy correlation \eqref{UofPpower} on varying material and geometric parameters relative to the baseline epithelial values of Table~\ref{tab:biological-baseline}.
 Varying $\eta_d, \eta_r$ over wide ranges has no appreciable effect. Varying relative cortical thickness $\tau$ and relative shear stiffness of the cytoplasm $\zeta$ does change $\alpha_2$ moderately, but only when the change also shifts the balance of energy dominance away from the cortex towards the cytoplasm, which occurs for smaller $\tau$ and larger $\zeta$, see figure (B) depicting the fraction of cytoplasmic (disk) energy. Even for extreme changes of material parameters the exponent does not approach the SVM value of 2.
    Shading width incorporates both fit errors and standard deviations across the polygonal population.}
 \label{fig:material}
\end{figure}
Similar analysis was applied to symmetric quadrilateral and octagonal shapes (blue and green curves in Fig.~\ref{fig:anharmonic}A, see SI for more detail). Edge numbers of 4 and 8 allow us to preserve the symmetry constraints of the shape space, while representing extreme cases: In confluent epithelial tissues, the population numbers of 4- and 8-sided cells are small ($\lesssim 3\%$ of the total each \cite{Gibson2006}), and cells of even larger or smaller edge number are virtually non-existent. As shown in Fig.~\ref{fig:anharmonic}A, the $\hat{U}(\mathcal Q)$ correlation of all polygon classes is very consistent, while the range of ${\mathcal Q}$ is extended (octagons can have smaller ${\mathcal Q}$ than hexagons).

In Fig.~\ref{fig:anharmonic}B, we show that this correlation is not an artifact of the symmetric shapes: A Monte-Carlo sampling of arbitrary hexagons (with no symmetry) yields near-identical $\hat{U}(\cal Q)$ results. Figure~\ref{fig:anharmonic}C averages the polygon topologies according to experimental population numbers in epithelia \cite{Gibson2006}, resulting in the black curve representing a perimeter energy functional in the spirit of a vertex model. However, the energy-perimeter correlation strongly differs from a quadratic function of perimeter,  contradicting the SVM. The yellow curve in Fig.~\ref{fig:anharmonic}C is a best-fit by least-squares relative errors, translating the variable ${\mathcal P}$ of the SVM \eqref{eq:svm} into ${\mathcal Q}$ (the functional dependence on ${\mathcal Q}$ is still very close to quadratic).
By contrast, a fit of the population curve with arbitrary power,
\begin{equation}
\langle \hat{U}\rangle = \alpha_1 {\mathcal Q}^{\alpha_2}
    \label{UofPpower}
\end{equation}
finds $\alpha_1\approx 27.3$ and an exponent $\alpha_2\approx 1.13$, much lower than the harmonic law of SVM. We note that $\alpha_1$ is not exactly $1/\mathcal{Q}_{\hexagon}^{\alpha_2}$ because of population averaging effects. This fit (dark red curve in Fig.~\ref{fig:anharmonic}C) is in excellent agreement with the continuum mechanics results.

The consequences of this finding are clearer when plotting the derivative $d\langle\hat{U}\rangle/d{\mathcal Q}$ as a measure of typical force on a vertex in a simulation \cite{sussman2017cellgpu}, see Fig.~\ref{fig:anharmonic}D: Between the lowest perimeter encountered (regular octagons, ${\mathcal Q}_{min}\approx 0.027$) and ${\mathcal Q}_{max}$, the harmonic elasticity law of the SVM predicts a change in vertex force by a factor of $>8$, while the subharmonic law based on continuum mechanics sees the force rise by less than $30\%$. Tensile forces in the edges are thus only weakly dependent on strain. Note that recent modeling of active cells \cite{kim2021embryonic,rizzi2026universal} has used vertex models with constant tension (making polygonal cells equivalent to 2D foam bubbles) with great success, while other literature has also questioned strong dependence of mechanics on deformation \cite{bera2026shape}. Because the present formalism is based on a complete continuum mechanics theory with defined moduli, the force scale has a direct dimensional equivalent, predicted as varying between 1.2\,nN and 1.5\,nN over the range of ${\mathcal Q}$. This agrees with a closed-form estimate
$F\simeq\frac{36\cdot6^{1/3}}{3\pi^2}\,
   G_r\,hR^2\tau^2\approx0.95\,\mathrm{nN}$ obtained from the present theory for deformation to a regular hexagon, taking into account the dominance of cortical bending, governed by $G_r$, see Materials and Methods for a derivation.
Measured local forces exerted by epithelial cells are consistent with that magnitude \cite{du2005force,roca2017quantifying}.

Viewing the perimeter of a cell as a rubber band with harmonic elasticity thus contradicts continuum elasticity.
An intuitive explanation for the relation \eqref{UofPpower} follows from considering individual Fourier modes (cf.\ Fig.~\ref{fig:shapes}A and \eqref{eq:prescribed-modal-perimeter} in Methods): Deforming the unstrained state solely in mode $n=0$ changes the perimeter linearly in the amplitude $f_0$, while the energy change $U_0$ is of course quadratic in $f_0$. Thus, $U_0\propto {\mathcal P}^2$ holds. For any other $n>0$ mode, the elastic energy change remains quadratic in $f_n$, while the perimeter change is to leading order quadratic as well (see Methods), yielding
$U_n\propto {\mathcal P}^1$. We have seen above that the deformation energy is dominated by $n>0$ modes because of the near-incompressibility of the soft material. Thus, their linear dependence on ${\mathcal P}$, and approximately on ${\mathcal Q}$, dominates the effective power law exponent $\alpha_2$.

One may be concerned about drawing conclusions from a linear elasticity formalism, as strains in the most deformed polygons of this study reach $\gtrsim 20\%$. We systematically investigated quadratic Hencky elasticity as a prototypical nonlinear formalism (see SI) and found that the conclusions drawn here are robust: cell elastic energies correlate strongly with perimeter, but never close to quadratically.

We emphasize that the correlation of $\langle\hat{U}\rangle$ and ${\mathcal Q}$ could be represented by functional dependencies other than pure power laws, and more accurately so if more parameters are used. The important distinction from the  SVM is that the resulting forces on vertices in a simulation are very different from what would be expected under the standard model.

\subsection*{Vertex Model dependence on material parameters}
The findings above were obtained with material and geometric parameter values taken from the literature on MDCK and other epithelia. Does anharmonic perimeter elasticity persist when parameters are varied? Our formalism suggests defining four dimensionless parameters, all small for the soft biological materials of concern here:
\begin{equation}
  \tau =\frac{\mathsf t}{\mathsf R},\quad
  \zeta =\frac{G_{\mathrm d}}{G_{\mathrm r}},\quad
  \eta_{\mathrm d} =\frac{G_{\mathrm d}}{K_{\mathrm d}},\quad
  \eta_{\mathrm r} =\frac{G_{\mathrm r}}{K_{\mathrm r}}\,,
\label{eq:material-ratios}
\end{equation}
with biological reference values of $\tau_{\rm ref}\approx 0.04, \zeta_{\rm ref}\approx 4.7\times 10^{-4}, \eta_{d,\rm ref}\approx 7.5\times 10^{-7}, \eta_{r,\rm ref}\approx 9.6\times 10^{-3}$ following from Table~\ref{tab:biological-baseline}.
Repeating the analysis above for different parameters, we focus on changes of the exponent $\alpha_2$ (the prefactor $\alpha_1$ of power-law fits stays close to $1/\mathcal{Q}_{\hexagon}^{\alpha_2}$ in all cases).
Figure~\ref{fig:material}A shows that, as long as the soft-matter properties $\eta_d,\eta_r\ll 1$ are respected, $\alpha_2$ remains quantitatively the same when their values are changed. Sweeping the remaining parameters $\tau$ and $\zeta$ by orders of magnitude, power laws \eqref{UofPpower} remain good fits (perimeter elasticity prevails as a robust concept). The value of $\alpha_2$, however, does change and rises if the
cortical thickness $\tau$ is decreased or the relative cytoplasmic shear stiffness $\zeta$ is increased: $\alpha_2$ is larger precisely when the dominance of cortical elastic energy is lost to the cytoplasm (Fig.~\ref{fig:material}B). Even for unrealistically thin or weak cortices, however, the exponent never reaches 2.
In realistic models of epithelia, the cortex does dominate elastic energy, and the perimeter dependence remains subharmonic with exponents robustly close to 1. %

\subsection*{Conclusions}
The present work aims at better insight into, and better justification of, the use of vertex models in tissue modeling. Rigorously computing elastic deformation energies for cells featuring cytoplasmic and cortical elastic continua, we draw three major conclusions regarding Standard Vertex Model approaches relating cell energies to perimeters and areas: (1) Due to the soft-matter character of the continua, areas are strongly confined to be close to their equilibrium values, rendering the area elasticity term unimportant. (2) Over a wide range of deformations, elastic energy $U$ does correlate strongly with perimeter changes measured as ${\mathcal P}$ or ${\mathcal Q}$, supporting vertex models with perimeter elasticity and explaining their success as consistent with the cortical dominance in stiffness. (3) The functional dependence of $U$ on perimeter measures is, however, not quadratic, but much closer to linear. This anharmonic elasticity, encountered over a wide range of continuum material parameters, challenges the accuracy of the SVM.

An immediate consequence of the anharmonic elasticity is a much greater uniformity of effective tensions in cell edges. This impacts not only the force magnitudes in vertex model simulations, but the conditions under which tissues behave rigidly or floppily: Loss of rigidity within vertex models \cite{Bi2015,park2015unjamming} can be understood as a competition between elastic tension and cell-cell adhesion (understood as a uniform lowering of edge tension), so that strong enough adhesion allows cells to relax into floppy states \cite{Staple10,Kim15}. With the anharmonic elasticity following from the continuum description, reaching this threshold requires much higher adhesion strength, i.e., tissues are less prone to loss of rigidity than assumed in the SVM \cite{jairo2025}.
Our results suggest that these conclusions remain valid when elastic continuum properties are changed, as long as the material obeys the near-incompressibility characteristic of soft matter. Adjustment of the effective anharmonic exponent may improve modeling for different types of confluent tissues. For biologically realistic parameters, cortical deformation dominates energy, with the largest contribution coming from hoop stress and strain. This confirms the intuition of perimeter-length elasticity, but highlights that the automatic assumption of quadratic dependence on perimeter change is not justified.

Vertex model simulations using the continuum mechanics-based elastic energy \eqref{UofPpower} are no more computationally expensive than the SVM, while the prefactors of energy and force terms are directly derived from material parameters. Future work will investigate how this change in the underlying elastic energy affects the use of morphological indicators like the shape index \cite{bi2014energy,Bi2015,park2015unjamming} or the quantitative use of geometric and statistical properties of the polygonal patterns \cite{Kim18,Kim19} in order to diagnose tissue mechanics and function from cell morphology \cite{lecuit2011force,heisenberg2013forces}.

\begin{materialsandmethods}

\subsection*{Epithelial cell dimensions}
While our modeling is not specific to a particular tissue, we take single-layer MDCK-II epithelial sheets \cite{hansson1986two} as a prototypical and well-studied system and choose our parameters from literature from mammalian epithelia of similar description.
Statistical analysis of images of single-layer confluent MDCK-II tissues show average areas $\bar{A}=\pi R^2$ with $R \approx 7.5\,\mu\mathrm{m}$ and heights of $hR\approx 7\,\mu\mathrm{m}$. These literature values \cite{Kaliman2016,Zehnder2015,Chmiel2022} are confirmed by our own analysis of images \cite{bill_data}.
No direct measurement of the lateral MDCK~II cortex thickness that is represented by the ring in our model is available, but analysis of basolateral cortex thickness in confluent MDCK~II cells \cite{Janshoff2021Basal},
fluorescence measurements in mitotic HeLa cells \cite{Clark2013}, and atomic-force microscopy of adherent HEK-293T cells \cite{Luchtefeld2020}, all support values of $\sim 0.15$\,$\mu\mathrm{m}$. Neighboring cells meet along a shared lateral cortex augmented by nanometer-scale membrane and cadherin layers, so that the disk--ring calculation represents the deforming cortical structure as one ring of thickness $\tau R\approx 0.30\,\mu\mathrm{m}$, with $\tau\approx 0.04$.

\subsection*{Material properties of epithelial cells}
Comprehensive, systematic measurements of elastic moduli in living cells differentiating cytoplasm and cortex are rare. The work by Guo et al.\ \cite{Guo2017} provides such values from a combination of optical-tweezer and magnetic forces on nano- and microbeads embedded in A7 cells derived from mammalian epithelia, as well as from osmotic pressure-volume measurements on the same cells. As we are interested in order-of-magnitude values of individual cells, and our analysis shows little variation over a wide range of material properties, we adopt the results of these measurements for definiteness. Moduli in \cite{Guo2017} are dependent on cell volume, as they change with externally applied osmotic compression. We seek to model tissues in the absence of such hydrostatic compression, so that we use values measured at physiological osmolarity in (Fig.~1(A) of Ref.~\cite{Guo2017}). This yields $G_{\mathrm d}\approx 2.4\,\mathrm{Pa}$ and $K_{\mathrm d}\approx 3.2\,\mathrm{MPa}$ from cytoplasmic measurements, as well as $G_{\mathrm r}\approx 5.1\,\mathrm{kPa}$ for the cortex.

No direct results of the cortical bulk modulus are available, so we infer a value from measurements of the (two dimensional) in-plane area extension modulus $K_A$ by indentation experiments on the apical cortex of MDCK-II cells in confluent monolayers \cite{cordes2020}, which yield an average of $K_A\approx 0.08\,\mathrm{N/m}$. This value contains contributions from both the membrane and the actin mesh~\cite{Janshoff2021Basal}, in keeping with our interpretation of the cortex as an effective shell structure. The cortex bulk modulus $K_r$ is then related to $K_A$ under the assumption of plane-strain isotropic linear elasticity  by~\cite{Hashin1972}
\begin{equation}
 K_{\mathrm r}=\frac{2K_A}{\tau R}-\frac{G_{\mathrm r}}{3}\,.
 \label{eq:cortical-modulus-conversion}
\end{equation}
The effective thickness here is $\tau R/2=0.15\,\mu\mathrm{m}$, as $K_A$ was measured on single apical cortices. We obtain $K_{\mathrm r}\approx 0.53\,\mathrm{MPa}$ and a Poisson ratio of $0.495$ for the ring material; the shear term in \eqref{eq:cortical-modulus-conversion} is nearly negligible.

\subsection*{Computing areas and perimeters of deformed cells}

The unstrained circular disk of our model is deformed into a perimeter outline (outer edge of the ring) described by the radial parametrization
$R+f(\theta)>0$.
The area and perimeter of this outline follow from polar geometry,
\begin{equation}
 \begin{aligned}
 A&=\frac12\int_0^{2\pi}[R+f(\theta)]^2\,d\theta,\\
 P&=\int_0^{2\pi}\sqrt{[R+f(\theta)]^2
 +[\partial_\theta f(\theta)]^2}\,d\theta.
 \end{aligned}
 \label{eq:prescribed-area-perimeter}
\end{equation}

For shapes symmetric about both coordinate axes, $f$ contains only even cosine modes of angular harmonic number $n$,
\begin{equation}
 f(\theta)=f_0+\sum_{n=2,4,\ldots}f_n\cos(n\theta).
 \label{eq:prescribed-displacement-fourier}
\end{equation}
where $f_0$ changes the radius uniformly and $f_2,f_4,\ldots$ describe the shape.
Substituting \eqref{eq:prescribed-displacement-fourier} gives the area exactly as
\begin{equation}
 \begin{aligned}
 A&=\pi(R+f_0)^2+\frac\pi2\sum_{n=2,4,\ldots}f_n^2,
 \end{aligned}
 \label{eq:prescribed-modal-area}
\end{equation}
while an expansion in small $n|f_n|/R$ yields the perimeter to second order in $f_n$,
\begin{equation}
 \begin{aligned}
 P&\simeq2\pi(R+f_0)
 +\frac\pi{2R}\sum_{n=2,4,\ldots}n^2f_n^2\,.
 \end{aligned}
 \label{eq:prescribed-modal-perimeter}
\end{equation}
For the straight-edged polygonal shape families in the present work, perimeters are computed directly from the vertex coordinates, while \eqref{eq:prescribed-modal-perimeter} makes different scaling of $P$ with $f_0$ and $f_n$ explicit.

\subsection*{Michell theory of elasticity}
Solutions to the equations of linear elasticity in polar coordinates can be found from solving a biharmonic equation for an Airy function \cite{Michell1899}, whose derivatives obtain stress components related to strain components by constitutive relations (see SI for details of the formalism). Using the shape decomposition \eqref{eq:prescribed-displacement-fourier}, the elastic energy can be evaluated for each mode separately to yield
\begin{equation}
 U=\frac12\left[k_0f_0^2
 +\sum_{n=2,4,\ldots}k_nf_n^2\right]\,,
 \label{eq:modal-energy-sum}
\end{equation}
where explicit calculation and integration of the energy densities $w_{d,r}$ obtains analytical relations for the spring constants $k_0, k_n$. They fulfill the implicit relations
\begin{equation}
 \begin{aligned}
 \biggl[\frac{k_0}{4\pi hR}+G_{\mathrm r}\biggr]^{-1}
 &=\frac{(1-\tau)^2}{K_{\mathrm d}+G_{\mathrm d}/3+G_{\mathrm r}}\\[3pt]
 &\quad+\frac{\tau(2-\tau)}{K_{\mathrm r}+4G_{\mathrm r}/3},
 \end{aligned}
 \label{eq:modal-spring-constant-zero}
\end{equation}
for $n=0$ and
\begin{equation}
 \begin{aligned}
 &\biggl[\tau(2-\tau)+\frac{1/G_{\mathrm r}+1/(K_{\mathrm r}+G_{\mathrm r}/3)}{2\pi hR(n^2-1)/k_n-1/G_{\mathrm r}}\biggr]^{-1}\\[4pt]
 &=(n+1)\biggl[n-(n+1)(1-\tau)^2\\
 &\phantom{=(n+1)\biggl[}+\frac{1/G_{\mathrm r}+1/G_{\mathrm d}+2/(K_{\mathrm r}+G_{\mathrm r}/3)}{(1-\tau)^{2n-2}(1/G_{\mathrm d}-1/G_{\mathrm r})}\biggr]^{-1}\\[4pt]
 &+(n-1)\biggl[n-(n-1)(1-\tau)^2+(1-\tau)^{2n+2}\\
 &\phantom{=(n+1)\biggl[}\times\frac{\begin{array}{@{}c@{}}1/G_{\mathrm r}+2/(K_{\mathrm r}+G_{\mathrm r}/3)\\-1/G_{\mathrm d}-2/(K_{\mathrm d}+G_{\mathrm d}/3)\end{array}}{1/G_{\mathrm r}+1/G_{\mathrm d}+2/(K_{\mathrm d}+G_{\mathrm d}/3)}\biggr]^{-1}
 \end{aligned}
 \label{eq:modal-spring-constant}
\end{equation}
for $n\ne 0$, respectively.
These expressions incorporate the contributions of both disk and ring (cytoplasm and cortex) for each mode.

For biologically relevant cases, these formulae simplify dramatically and allow for an estimate of the characteristic force scale in a vertex model based on continuum theory. For definiteness, we estimate this scale for a regular hexagon, by $F\sim U_{\hexagon}/(2\pi R {\cal Q}_{\hexagon})$, the deformation energy divided by the perimeter change. The Fourier decomposition of a regular hexagon shape only contains shape amplitudes with $n=6k$, and $|f_{6k}|\simeq |f_6|/k^2$. The perimeter length \eqref{eq:prescribed-modal-perimeter} then gives
$ \mathcal{Q}_{\hexagon}\simeq\sum_nn^2f_n^2/(4R^2)=(3\pi^2/2)\,f_6^2/R^2$. Contributions of mode $n$ to energy, on the other hand, decay steeply for $n>n_c=6^{1/3}/\tau\approx45$, because for these high modes the material does not deform like a thin ring any more. For the relevant modes $n\leq n_c$, and for the material parameters appropriate for tissue ($\eta_r,\eta_d,\zeta\ll 1$) the stiffnesses \eqref{eq:modal-spring-constant} simplify to
\begin{equation}
k_n \simeq \frac{\pi}{3} h R G_r \tau^3 n^4,
\label{eq:knsimple}
\end{equation}
and their growth compensates for the decay of the $f_n^2$ amplitudes, so that each $U_n$ contains roughly the same energy (cf.\ SI). Multiplying $U_6=\frac{1}{2}k_6f_6^2$ by the number $n_c/6$ of relevant non-zero modes obtains
\begin{equation}
 U_{\hexagon}\simeq\frac{n_c}{6}\,\frac{k_6R^2\mathcal{Q}_{\hexagon}}{3\pi^2}.
 \label{eq:hexagon-energy-scale}
\end{equation}
With \eqref{eq:knsimple} for $n=6$, dividing this result by $2\pi R {\cal Q}_{\hexagon}$ yields the force scale given in the main text,
\begin{equation}
 F\simeq\frac{36\cdot6^{1/3}}{3\pi^2}\,
   G_r\,hR^2\tau^2\approx0.95\,\mathrm{nN}.
 \label{eq:hexagon-energy-force}
\end{equation}

\subsection*{Averaging across the cell shape space}

Our main results relate elastic energies to shape excess $\mathcal Q$. Within the hexagonal shapes, for example, the shapes with $\mathcal Q=q$ for given $q$ form a level set whose members differ in shape and energy (Fig.~\ref{fig:shapes}B). Any observable $X$, such as $U$ or $f_n^2$, is averaged over the level set, with each shape evaluated at its own area $A_{eq}$. Because $\mathcal Q$ depends on shape alone, its level sets are fixed by geometry and do not change with the material, which enters only through $X$.
The average of $X$ at given $q$ then follows from integration over the contour measures $ds$ in $\boldsymbol v_2$-space, weighted by the gradient (variation perpendicular to contour lines) $\lVert\nabla\mathcal Q\rVert$, i.e. \cite{Federer1969}
\begin{equation}
 \langle X \rangle (q)
 =\frac{\displaystyle\int_{\mathcal Q(\boldsymbol v_2)=q}
 \frac{X\,ds}{2v_{2,y}\lVert\nabla_{\boldsymbol v_2}\mathcal Q\rVert}}
 {\displaystyle\int_{\mathcal Q(\boldsymbol v_2)=q}
 \frac{ds}{2v_{2,y}\lVert\nabla_{\boldsymbol v_2}\mathcal Q\rVert}}.
 \label{eq:relaxed-conditioned-expectation}
\end{equation}
The integrals run over the level set contour, whether it is simply connected or not, and the factor $(1/2v_{2,y})$ is the rescaling that the area constraint induces on the chart in $\boldsymbol v_2$-space.
Equation~(\ref{eq:relaxed-conditioned-expectation}) obtains averages like $\langle \hat{U} \rangle$ in Fig.~\ref{fig:anharmonic}. The conditional variance at given $q$ follows as usual,
\begin{equation}
 \mathbb{V}[X\mid\mathcal Q=q]
 = \langle X^2\rangle (q) - \left[ \langle X \rangle (q) \right]^2\,.
 \label{eq:relaxed-conditioned-spread}
\end{equation}
Its square root is the standard deviation drawn as shaded bands in the figures.

\subsection*{Declaration of AI use}

Claude (Anthropic) and Codex (OpenAI), and to a minor extent other AI assistants, were used for coding support, literature search, checking derivations, and editing the manuscript. All AI-generated suggestions were reviewed and verified by the authors, who take full responsibility for the scientific content and conclusions of the study.

\end{materialsandmethods}

\vspace*{-0.2cm}

\begin{acknowledgments}
The authors are indebted to William Brieher, Vivian Tang, and Ran Yang for sharing experimental data on epithelial tissue morphology. We are grateful for stimulating discussions with William Brieher, Sangwoo Kim, and Ran Yang.
\end{acknowledgments}

\phantomsection\label{mainlastpage}

\clearpage
\onecolumn
\pagestyle{supplement}
\setcounter{page}{1}
\renewcommand{\thepage}{S\arabic{page}}
\renewcommand{\rmdefault}{lmr}
\renewcommand{\sfdefault}{lmss}
\normalfont
\makeatletter
\let\eqref\amseqref
\renewcommand\tagform@[1]{\maketag@@@{(\ignorespaces#1\unskip\@@italiccorr)}}
\makeatother
\setcounter{equation}{0}
\setcounter{figure}{0}
\renewcommand{\theHequation}{si.\arabic{equation}}
\renewcommand{\theHfigure}{si.\arabic{figure}}
\renewcommand{\theHsubsection}{si.\arabic{subsection}}
\renewcommand{\topfraction}{0.9}
\renewcommand{\bottomfraction}{0.8}
\renewcommand{\textfraction}{0.1}
\renewcommand{\floatpagefraction}{0.7}
\setcounter{topnumber}{1}\setcounter{bottomnumber}{2}\setcounter{totalnumber}{4}
\makeatletter\setlength{\@fptop}{0pt}\makeatother
\captionsetup{font=small,labelfont={},textfont={},labelsep=period}
\hypersetup{allcolors=black}
\setcitestyle{square}
\renewcommand{\bibfont}{\normalfont\small}
\renewcommand{\bibnumfmt}[1]{[#1]}
\setcounter{secnumdepth}{3}
\renewcommand{\thesubsection}{S\arabic{subsection}}
\renewcommand{\thesubsubsection}{\thesubsection.\arabic{subsubsection}}
\titleformat{\subsection}{\normalfont\normalsize\bfseries\raggedright}{\thesubsection.}{1em}{#1}
\titleformat{\subsubsection}{\normalfont\normalsize\bfseries\raggedright}{\thesubsubsection}{1em}{#1}
\titlespacing*{\subsection}{0pt}{.8cm plus1ex minus .2ex}{.4cm}
\titlespacing*{\subsubsection}{0pt}{.6cm plus1ex minus .2ex}{.3cm}

\begin{center}
  {\large\bfseries Supporting Information to\\ A Tissue Vertex Model grounded in Continuum Mechanics\par}
  \vskip1em
  {Jairo M. Rojas,$^{1,\,2}$ Mayisha Z. Nakib,$^{1,\,2}$ and Sascha Hilgenfeldt$^{1,\,2}$\par}
  \vskip.5em
  {\small\itshape
  $^{1}$Department of Physics, The Grainger College of Engineering,\\
  University of Illinois, Urbana-Champaign, Illinois 61801, USA\\
  $^{2}$Mechanical Science and Engineering, The Grainger College of Engineering,\\
  University of Illinois, Urbana-Champaign, Illinois 61801, USA\par}
\end{center}
\vskip1em

\renewcommand{\figurename}{Supplementary Fig.}
\setcounter{figure}{0}
\renewcommand{\thefigure}{S\arabic{figure}}

\subsection{Michell elastic theory}
The Michell theory of linear elasticity \cite{si:Michell1899,si:hulbert1963numerical,si:meleshko2003selected} can be formulated in terms of an Airy function
$\Phi_s$ fulfilling a biharmonic equation; the subscript $s=d,r$ differentiates between disk (cytoplasm) and ring (cortex) in the following. For the present plane-strain formalism, the relevant derivative is the in-plane Laplacian; in polar coordinates,
\begin{equation}
 \nabla_\perp^2\equiv\frac1{R^2}\left(
 \partial_\rho^2+\frac1\rho\partial_\rho
 +\frac1{\rho^2}\partial_\theta^2\right),
 \label{si:eq:in-plane-laplacian}
\end{equation}
and the biharmonic equation reads
\begin{equation}
 \nabla_\perp^2\bigl(\nabla_\perp^2\Phi_s\bigr)
 =\nabla_\perp^4\Phi_s=0.
 \label{si:eq:biharmonic-compatibility}
\end{equation}
Stress components are obtained by derivatives of $\Phi_s$; in polar coordinates we have
\begin{equation}
 \begin{aligned}
 \sigma_{\rho\rho,s}
 &=\frac{1}{R^2}\left(\frac{\partial_\rho\Phi_s}{\rho}
 +\frac{\partial_\theta^2\Phi_s}{\rho^2}\right),\\
 \sigma_{\theta\theta,s}&=\frac{\partial_\rho^2\Phi_s}{R^2},\\
 \sigma_{\rho\theta,s}
 &=-\frac{1}{R^2}\partial_\rho\left(\frac{\partial_\theta\Phi_s}{\rho}\right).
 \end{aligned}
 \label{si:eq:airy-stress-function}
\end{equation}
These stress components automatically fulfill mechanical equilibrium, i.e.,
\begin{equation}
 \nabla_\perp\cdot\boldsymbol\sigma_s=\mathbf0.
 \label{si:eq:in-plane-equilibrium}
\end{equation}
The plane-strain constitutive relation between components of stress $\sigma_{ik,s}$ and strain $\epsilon_{ik,s}$ is
\begingroup
\setlength{\arraycolsep}{3pt}
\begin{equation}
\begin{bmatrix}
 \sigma_{\rho\rho,s}\\
 \sigma_{\theta\theta,s}\\
 \sigma_{\rho\theta,s}
\end{bmatrix}
=K_{\mathrm r}\eta_{\mathrm r}\zeta^{\delta_{s\mathrm d}}
\begin{bmatrix}
 \frac{1}{\eta_s}+\frac43 & \frac{1}{\eta_s}-\frac23 & 0\\
 \frac{1}{\eta_s}-\frac23 & \frac{1}{\eta_s}+\frac43 & 0\\
 0 & 0 & 2
\end{bmatrix}
\begin{bmatrix}
 \epsilon_{\rho\rho,s}\\
 \epsilon_{\theta\theta,s}\\
 \epsilon_{\rho\theta,s}
\end{bmatrix},
\label{si:eq:plane-strain-law}
\end{equation}
\endgroup
where we use the shorthand $\zeta^{\delta_{sd}}$ with the Kronecker symbol $\delta_{sd}$: for the disk material, this replaces the modulus scale $K_r$ by $K_d$.

By definition, the strain components in the plane are related to displacements $u_j$ by
\begin{equation}
 \begin{gathered}
  \epsilon_{\rho\rho}=\frac{\partial_\rho u_\rho}{R},
  \qquad
  \epsilon_{\theta\theta}
  =\frac{u_\rho}{R\rho}
   +\frac{\partial_\theta u_\theta}{R\rho},\\
\epsilon_{\rho\theta}
  =\frac{\partial_\rho u_\theta}{2R}
   -\frac{u_\theta}{2R\rho}
   +\frac{\partial_\theta u_\rho}{2R\rho},
 \end{gathered}
 \label{si:eq:infinitesimal-strain-3d}
\end{equation}
To ensure that the strains result from a consistent displacement field, they need to fulfill the (plane-strain) compatibility condition~\cite{si:Timoshenko1970}
\begin{equation}
 \begin{aligned} &\partial_\rho\bigl(\rho^2\partial_\rho\epsilon_{\theta\theta,s}\bigr)
 -\rho\partial_\rho\epsilon_{\rho\rho,s}\\
 &\qquad +\partial_\theta^2\epsilon_{\rho\rho,s}
-2\partial_\rho\bigl(\rho\partial_\theta\epsilon_{\rho\theta,s}\bigr)=0.
 \end{aligned}
 \label{si:eq:polar-strain-compatibility}
\end{equation}

\subsubsection{Boundary conditions}
\label{si:sec:boundary-conditions}

We prescribe a cellular outline with radial departure $f(\theta)$ from the reference radius $R$ by the Fourier modes
\begin{equation}
 f(\theta)=f_0+\sum_{n=2,4,\ldots}f_n\cos(n\theta),
 \label{si:eq:prescribed-displacement-fourier}
\end{equation}
describing the radial displacement of the rim of the ring.
The rim carries no shear stress, because neighboring cells are taken to exert no tangential force on each other, so it slides freely along the outline,
\begin{equation}
 \left.u_{\rho,\mathrm r}\right|_{\rho=1}=f(\theta),
 \qquad
 \left.\sigma_{\rho\theta,\mathrm r}\right|_{\rho=1}=0.
 \label{si:eq:outer-tractions}
\end{equation}

At the interface between disk and ring the two materials are bonded, so both displacement components are continuous,
\begin{equation}
 \begin{gathered}
  \left.u_{\rho,\mathrm d}\right|_{\rho=1-\tau}
  =\left.u_{\rho,\mathrm r}\right|_{\rho=1-\tau},\\
  \left.u_{\theta,\mathrm d}\right|_{\rho=1-\tau}
  =\left.u_{\theta,\mathrm r}\right|_{\rho=1-\tau},
 \end{gathered}
 \label{si:eq:interface-conditions}
\end{equation}
and the interface transmits radial and tangential tractions,
\begin{equation}
 \begin{gathered}
  \left.\sigma_{\rho\rho,\mathrm d}\right|_{\rho=1-\tau}
  =\left.\sigma_{\rho\rho,\mathrm r}\right|_{\rho=1-\tau},\\
  \left.\sigma_{\rho\theta,\mathrm d}\right|_{\rho=1-\tau}
  =\left.\sigma_{\rho\theta,\mathrm r}\right|_{\rho=1-\tau}.
 \end{gathered}
 \label{si:eq:interface-stress-conditions}
\end{equation}

\subsubsection{Deformation energy computation}
\label{si:sec:energy-computation}
The decomposition into orthogonal Fourier modes translates to strains, stresses, and the Airy function. For each mode, the components of these variables are proportional to $f_n$. Decomposition of the Airy function
\begin{equation}
 \Phi_s=\sum_{n=0,2,\ldots}\Phi_{s,n}(\rho)\cos(n\theta),
 \label{si:eq:modal-fields}
\end{equation}
results in the following admissible solutions of the biharmonic equation:
\begin{equation}
 \begin{aligned}
 \Phi_{\mathrm d,0}&=K_{\mathrm r}Rf_0A_{\mathrm d,0}\rho^2,\\
 \Phi_{\mathrm r,0}&=K_{\mathrm r}Rf_0\bigl(A_{\mathrm r,0}\rho^2
 +C_{\mathrm r,0}\log\rho\bigr).
 \end{aligned}
 \label{si:eq:airy-mode-zero}
\end{equation}
for the $n=0$ mode and
\begin{equation}
 \begin{aligned}
 \Phi_{\mathrm d,n}&=K_{\mathrm r}Rf_n\bigl(A_{\mathrm d,n}\rho^n+C_{\mathrm d,n}\rho^{n+2}\bigr),\\
 \Phi_{\mathrm r,n}&=K_{\mathrm r}Rf_n\bigl(A_{\mathrm r,n}\rho^n+B_{\mathrm r,n}\rho^{-n}\\
 &\quad+C_{\mathrm r,n}\rho^{n+2}+D_{\mathrm r,n}\rho^{2-n}\bigr).
 \end{aligned}
 \label{si:eq:airy-mode-n}
\end{equation}
for $n\ne 0$. The coefficients $A_{s,n},B_{s,n},C_{s,n},D_{s,n}$ are known analytically, namely
\begin{equation}
 \begin{aligned}
 A_{\mathrm d,0}
 &=\frac{\eta_{\mathrm r}\zeta(\eta_{\mathrm d}+3)(4\eta_{\mathrm r}+3)}{3\mathfrak B_0},\\
 A_{\mathrm r,0}
 &=\frac{\eta_{\mathrm r}(\eta_{\mathrm r}+3)[\zeta(\eta_{\mathrm d}+3)+3\eta_{\mathrm d}]}{3\mathfrak B_0},\\
 C_{\mathrm r,0}
 &=\frac{2\eta_{\mathrm r}(1-\tau)^2}{\mathfrak B_0}\\
 &\quad\times[\eta_{\mathrm r}\zeta(\eta_{\mathrm d}+3)-\eta_{\mathrm d}(\eta_{\mathrm r}+3)].
 \end{aligned}
 \label{si:eq:mode-zero-coefficients}
\end{equation}
with
\begin{equation}
 \begin{aligned}
 \mathfrak B_0&=\eta_{\mathrm r}\zeta(\eta_{\mathrm d}+3)[1-(1-\tau)^2]\\
 &\quad+\eta_{\mathrm d}[3\eta_{\mathrm r}+(3+\eta_{\mathrm r})(1-\tau)^2].
 \end{aligned}
 \label{si:eq:B0}
\end{equation}
for $n=0$ and
\begin{equation}
 (A_{\mathrm r,n},B_{\mathrm r,n},C_{\mathrm r,n},D_{\mathrm r,n})
 =\frac{1}{\mathfrak B_n}
 (\mathfrak C_{3,n},\mathfrak C_{4,n},\mathfrak C_{1,n},\mathfrak C_{2,n}),
 \label{si:eq:cancellation-solution}
\end{equation}
with
\begin{equation}
 \mathfrak B_n=\frac1{2\eta_{\mathrm r}}
 \left[-n\mathfrak C_{3,n}+n\mathfrak C_{4,n}
 -\left(n-\frac{6\eta_{\mathrm r}}{3+\eta_{\mathrm r}}\right)\mathfrak C_{1,n}
 +\left(n+\frac{6\eta_{\mathrm r}}{3+\eta_{\mathrm r}}\right)\mathfrak C_{2,n}\right],
 \label{si:eq:prescribed-modal-denominator}
\end{equation}
\begin{align}
 \mathfrak C_{1,n}
 &=(1-n)n(3+\eta_{\mathrm r})\mathfrak a
 \Bigl[\mathfrak d(1-\tau)^2+(3+\eta_{\mathrm r})(\zeta-1)(1-\tau)^{2n}
 [1+(1+n)((1-\tau)^2-1)]\Bigr],
 \label{si:eq:modal-numerator-0}\\
 \mathfrak C_{2,n}
 &=n(1+n)(3+\eta_{\mathrm r})(\zeta-1)(1-\tau)^{2n}
 \Bigl[\mathfrak b(1-\tau)^{2n+2}-(3+\eta_{\mathrm r})\mathfrak a
 [-1+(n-1)((1-\tau)^2-1)]\Bigr],
 \label{si:eq:modal-numerator-1}\\
 \mathfrak C_{3,n}
 &=(1+n)(1-\tau)^2\Bigl\{
 n(3+\eta_{\mathrm r})\mathfrak a[\mathfrak d+(3+\eta_{\mathrm r})(\zeta-1)(1-\tau)^{2n}]
 +12(3+4\eta_{\mathrm r})\zeta\mathfrak c(1-\tau)^{2n+2}\notag\\
 &\qquad+n^2(3+\eta_{\mathrm r})^2(\zeta-1)\mathfrak a
 [(1-\tau)^2-1](1-\tau)^{2n}\Bigr\},
 \label{si:eq:modal-numerator-2}\\
 \mathfrak C_{4,n}
 &=(1-n)(1-\tau)^{2n+2}\Bigl\{
 n(3+\eta_{\mathrm r})(\zeta-1)[(3+\eta_{\mathrm r})\mathfrak a+\mathfrak b(1-\tau)^{2n}]
 -12(3+4\eta_{\mathrm r})\zeta\mathfrak c(1-\tau)^2\notag\\
 &\qquad-n^2(3+\eta_{\mathrm r})^2(\zeta-1)\mathfrak a[(1-\tau)^2-1]\Bigr\}
 \label{si:eq:modal-numerator-3}
\end{align}
and
\begin{equation}
 \begin{aligned}
 C_{\mathrm d,n}
 &=\frac{2(3+4\eta_{\mathrm r})\zeta(3+\eta_{\mathrm d})}
 {(3+\eta_{\mathrm r})\mathfrak a}\,C_{\mathrm r,n},\\
 A_{\mathrm d,n}
 &=A_{\mathrm r,n}-\frac{n+1}{n-1}B_{\mathrm r,n}(1-\tau)^{-2n}
 +\frac{n+1}{n-1}(C_{\mathrm r,n}-C_{\mathrm d,n})(1-\tau)^2
 -D_{\mathrm r,n}(1-\tau)^{2-2n}.
 \end{aligned}
 \label{si:eq:numerators-explicit}
\end{equation}
using
\begin{equation}
 \begin{aligned}
 \mathfrak a&=3+7\eta_{\mathrm d}+\zeta(3+\eta_{\mathrm d}),
 &\mathfrak c&=3\eta_{\mathrm r}\zeta-3\eta_{\mathrm d}+\eta_{\mathrm r}(\zeta-1)\eta_{\mathrm d},\\
 \mathfrak b&=9(\zeta-1)+3(\zeta-7)\eta_{\mathrm d}
 +\eta_{\mathrm r}[-3-7\eta_{\mathrm d}+7\zeta(3+\eta_{\mathrm d})],
 &\mathfrak d&=3+\eta_{\mathrm r}+3\zeta+7\eta_{\mathrm r}\zeta.
 \end{aligned}
 \label{si:eq:auxiliary-constants}
\end{equation}
for $n\ne 0$.

Each mode's stress and strain can then be computed and combined to yield the energy density of mode $n$,
\begin{equation}
 \begin{aligned}
 w_s^{(n)}
 &=\frac12\bigl[
 \sigma_{\rho\rho,s,n}\epsilon_{\rho\rho,s,n}
 +\sigma_{\theta\theta,s,n}\epsilon_{\theta\theta,s,n}
 \bigr]\cos^2(n\theta)\\
 &\quad+\sigma_{\rho\theta,s,n}\epsilon_{\rho\theta,s,n}
 \sin^2(n\theta).
 \end{aligned}
 \label{si:eq:single-mode-energy-density}
\end{equation}
As the energy density is quadratic in the relative amplitude, we can isolate the part independent of $f_n$ by writing $w_s^{(n)}=(f_n/R)^2\breve w_s^{(n)}$. The total deformation energy is then a sum over modes written as
\begin{equation}
 U=\sum_{n=0,2,\ldots}f_n^2\,hR\int_0^{2\pi}\Bigl[
 \int_0^{1-\tau}\breve w_{\mathrm d}^{(n)}\rho\,d\rho
 +\int_{1-\tau}^{1}\breve w_{\mathrm r}^{(n)}\rho\,d\rho
 \Bigr]d\theta\,,
 \label{si:eq:energy-unit-deformation}
\end{equation}
where mixed terms vanish by orthogonality.
The integrals can be evaluated analytically and the energy cast in the form of individual harmonic oscillator contributions,
\begin{equation}
 U=\frac12\left[k_0f_0^2
 +\sum_{n=2,4,\ldots}k_nf_n^2\right].
 \label{si:eq:modal-energy-sum}
\end{equation}
with spring constants $k_0$ and $k_n$ as given in the Materials and Methods section. We restrict the modes taken into account to $n\le n_{\max}\approx\pi/\tau$, as higher modes describe strong non-uniformity across the thickness of the thin cortex. For the biological reference parameters of Table 1 in the main text, $n_{\max}=78$, and we use this value throughout, also when the material parameters are varied.
\begin{figure}[tbp]
  \centering
  \includegraphics[width=0.7\linewidth]{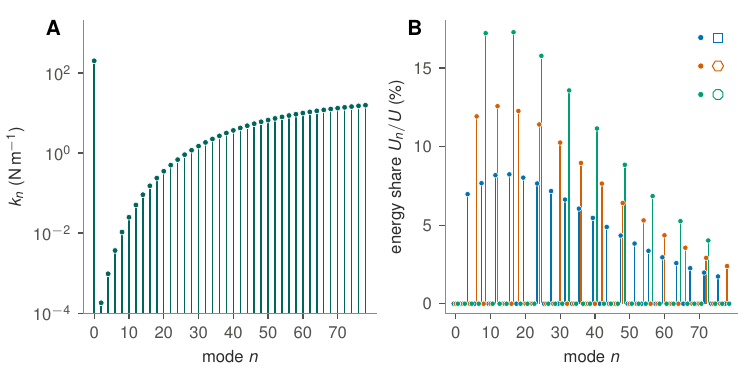} %
  \caption{(A) Spectrum of even-mode spring constants $k_0,k_n$ for the material parameters of Table 1 of the main text. (B) Elastic energy contributions $\frac{1}{2} k_nf_n^2/U$ of the modes for the displacements $f_n$ deforming a circle into a regular square (blue), hexagon (red), or octagon (green).}
  \label{si:fig:knun}
\end{figure}
Figure~\ref{si:fig:knun}A depicts these spring constants on a logarithmic scale, showing that $k_0$ (dominated by area changes) is much larger than the shape mode stiffnesses. When multiplied by the $f_n^2$ amplitudes of deformations, for example into regular polygons (Fig.~\ref{si:fig:knun}B), the energy contributions $U_n$ tend to peak at moderate $n$ and show a universal decay for large $n$.

\subsection{Parametrization of polygonal shape families}
\begin{figure}[tp]
  \centering
  \includegraphics[width=0.48\linewidth]{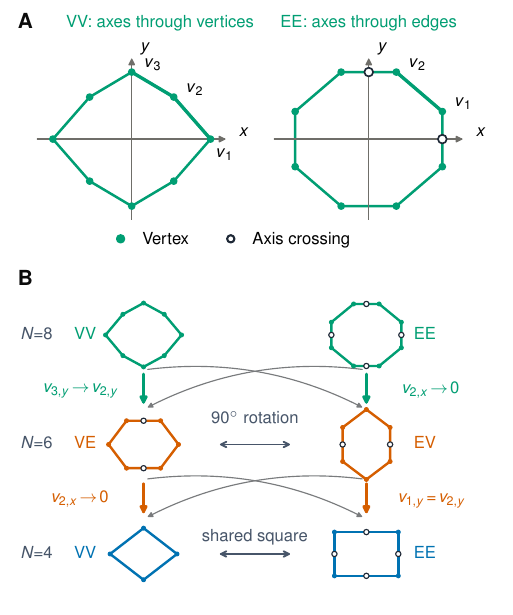} %
  \caption{Systematic parametrization of even-sided symmetric polygons. Shape families are classified by outlines intersecting axes at an edge (E) or vertex (V). Quadrilateral polygons have one degree of freedom fewer than the hexagon parametrization, while octagons have one more.}
  \label{si:fig:shapefamilies}
\end{figure}

In Fig.~4A, the main text demonstrates consistency of the interdependence of elastic energy and perimeter measures between polygonal shape families of different topology (edge number). As in the case of hexagons, we systematically explore shapes that preserve symmetry with respect to the $x$- and $y$-axes. It is convenient to classify the polygons by whether they intersect these two axes perpendicularly at the center of an edge (E), or whether they have a vertex lying on the axis (V), see Fig.~\ref{si:fig:shapefamilies}A. For hexagons, EV and VE shapes are equivalent and can thus be parametrized by a single continuous change in the coordinates of the $v_2$ vertex (Fig.~\ref{si:fig:shapefamilies}B). Quadrilaterals are parametrized as two families of degenerate hexagons, VV ($v_{2,x}=0$) and EE ($v_{a,y}=v_{2,y}$), respectively. Octagons likewise have two distinct shape family parametrizations with one additional vertex degree of freedom compared to the hexagons. In VV, hexagons are degenerate octagons with $v_{2,y}=v_{3,y}$, while the EE family yields hexagons when $v_{2,x}=0$. All of the vertex coordinates are scaled relative to $R$, so that the shape families remain the same upon changing the area of the polygon. After averaging over the respective parameter spaces, the differences between families of polygons with different topologies are small (Fig.~4A of the main text); in principle, a more detailed analysis could lead to the formulation of a vertex model with topology-dependent parameters, but in the interest of maintaining a simple energy-perimeter law, we use a single power-law fit to $\hat{U}({\mathcal Q})$ for the population mean averaged by probabilities of topologies in typical epithelia \cite{si:Gibson2006}.

\subsection{Perimeter excess and shape excess measures}
\begin{figure}[tbp]
  \centering
  \includegraphics[width=0.48\linewidth]{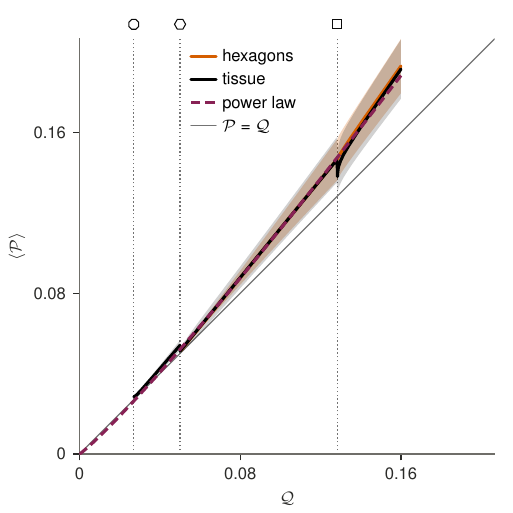} %
  \caption{Computation of perimeter excess ${\cal P}$, averaged over all shapes of a given ${\mathcal Q}$. The functional relationship between the two perimeter measures stays close to the identity line (solid), while the standard deviation (shaded) is small.}
  \label{si:fig:perimetermeasures}
\end{figure}

In the standard vertex model, area $A$ and perimeter $P$ enter relative to the unstrained area $A_0$ and the corresponding circular perimeter $P_0=(4\pi A_0)^{1/2}$. This poses two problems: (i) When analyzing experimental data, the only accessible areas are those of strained cells, so that $A_0$ is unknown (without detailed knowledge of mechanical parameters); (ii) The perimeter excess measure ${\cal P}=(P-P_0)/P_0$ is evaluated relative to $P_0$, which is not only unknown, but tied to the area $A_0$, so that ${\cal P}$ does not represent a pure shape change at constant area. For these reasons, we parametrize our theory with the shape excess ${\mathcal Q}=(P-P_c)/P_c$, a measure relative to the perimeter $P_c$ of a circular cell of current area $A$, $P_c=(4\pi A)^{1/2}$. In order to provide a fair comparison between SVM results and continuum modeling results, we have to take the differences between ${\cal P}$ and ${\mathcal Q}$ into account and plot both vertex model laws with respect to ${\mathcal Q}$. Figure~\ref{si:fig:perimetermeasures} shows that the difference between ${\cal P}$ and ${\mathcal Q}$ is slight and only deviates appreciably from identity for the highest values of ${\mathcal Q}$ taken into account in the present work. As a result, the SVM power law ${\hat U}\propto {\cal P}^2$ translates into ${\hat U}\propto {\mathcal Q}^\beta$ with a best-fit exponent of $\beta\approx 2.21$. It is this fit presented in Fig.~4C of the main text.

\subsection{Fitting data to vertex models}
In order to fit simple vertex model laws ${\hat U}({\mathcal Q})$ to the data of the continuum mechanics computations, one can determine quality of "best fit" in different ways. In all cases, we take the computational relationship for the population-averaged polygons (black line in Fig.~4C of the main text) as the reference data to be fitted. We restrict ourselves to pure power laws in ${\cal P}({\mathcal Q})$ for SVM (with exponent 2)  and with arbitrary exponent $\alpha_2$ in ${\mathcal Q}$ for the continuum model, respectively. Both of these functional dependencies automatically fulfill ${\hat U}_0$=0, an assumption underlying both models, even though the data to be fitted do not include ${\mathcal Q}=0$ (see dashed lines in Fig~4C of the main text).

\begin{figure}[tbp]
  \centering
   \includegraphics[width=0.7\linewidth]{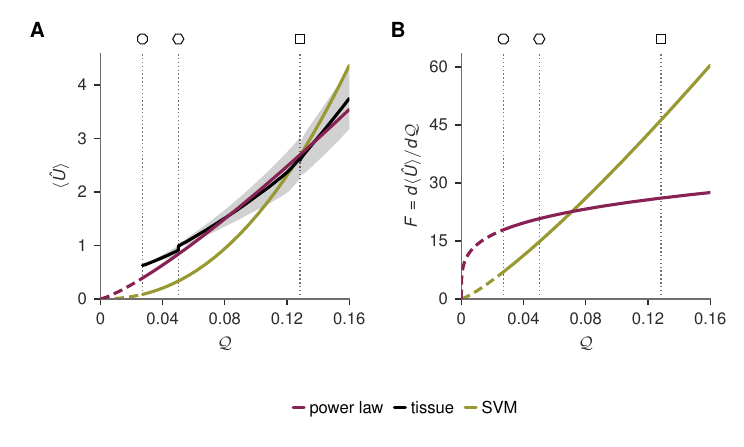} %
  \caption{(A) Fits minimizing least squares of absolute, rather than relative errors of averaged energies from the population-averaged computations (black line, identical to the one in Fig.~4C of the main text). Yellow is the SVM, dark red the fit with arbitrary exponent $\alpha_2$, with much better fit quality and close to a linear relation. (B) Derivatives of the fit curves in (A) representing vertex force scales inferred in either model, again showing the much weaker dependence of forces on strains in the continuum mechanics-informed model (dark red).
  }
  \label{si:fig:fiterrors}
\end{figure}

To determine best-fit parameters, least-squares errors between fits and data lines are employed, where the choice is to minimize relative or absolute errors. We present fits minimizing relative errors $|(\hat{U}_{\mathrm fit}-\hat{U}_{\mathrm data})/\hat{U}_{\mathrm data}|$ in the main text, as we expect less-deformed, more regular cell shapes with smaller ${\mathcal Q}$ to be more common in practically encountered data, whose smaller absolute $\hat{U}$ values lead to smaller weights in absolute errors $|(\hat{U}_{\mathrm fit}-\hat{U}_{\mathrm data})|$. We present a least-squares fit minimizing absolute errors for comparison in Fig.~\ref{si:fig:fiterrors}A, which, as expected, stays closer to the data at large ${\mathcal Q}$, but misses the computational values by a large factor for smaller ${\mathcal Q}$. The overall conclusions are unchanged: the power-law fit with general power (for absolute-error least-squares we have $\alpha_2\approx 1.24$) leads to much better fit quality than SVM and pinpoints an exponent closer to a linear relationship between $\hat{U}$ and ${\mathcal Q}$.
    Quantitatively, our general power law fit has a mean error of $4.3\%$ when using relative errors (main text) and $6.0\%$ when using absolute errors (this section), while the SVM best-fit errors are $38.5\%$ and $72\%$, respectively.
Derivatives of the fit curves in Fig.~\ref{si:fig:fiterrors}A, as measures of typical vertex forces following from either model, confirm the conclusions of Fig.~4D of the main text (Fig.~\ref{si:fig:fiterrors}B): The continuum mechanics-informed model (dark red) results in
much weaker dependence of forces on perimeter deformation than the SVM (yellow).

\subsection{Equilibrium areas of cells}
\begin{figure}[tbp]
  \centering
  \includegraphics[width=0.48\linewidth]{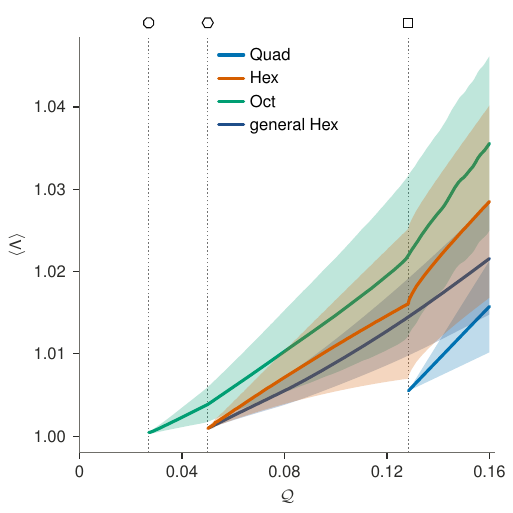} %
  \caption{Average scale factor $\Lambda$ between cells of equilibrium area and cells of nominal unstrained area for different polygon edge numbers. The deviations from the value 1 are systematically growing with ${\mathcal Q}$ and dependent on cell topology, but stay small.}
  \label{si:fig:cellareas}
\end{figure}

Even though our continuum mechanics model takes both area and shape changes of cells fully into account, it ultimately decouples them because of the strong constraint on area following from the overwhelmingly large dependence of elastic energy on $A$ (Fig.~3A of the main text). Equilibrium areas $A_{eq}$ minimizing the contribution of mode $n=0$ to energy are individually dependent on the shape of each cell, and thus vary over a certain range. They are not equal to the area $A_0$ nominally assigned to unstrained cells. Note that in analysis of experimental data and images, cell outlines only define an instantaneous area $A$, which in the model can then be assumed as equal to $A_{eq}$. The underlying distribution of unstrained areas $A_0$ in e.g.\ tissue samples is different from the $A_{eq}$ distribution. We quantify this by a parameter $\Lambda^2=A_{eq}/A_0$ ($\Lambda$ is a length scaling factor). Figure~\ref{si:fig:cellareas} shows that this  scale factor $\Lambda$ stays within a few percent of the value 1, varies systematically  with ${\mathcal Q}$ and cell topology, and is minimal for maximally symmetric shapes. In practical terms, polydispersity measures of tissues will not be greatly affected by the difference between areas accessible by visual analysis ($A_{eq}$) and those inferred from mechanics ($A_0$).

\subsection{Nonlinear Elasticity}
The deformations of continuum polygons representing cells in the present work are not small in the sense of traditional continuum elasticity, as is appropriate for soft-matter materials whose range of relevant strains tends to be large. A large body of fundamental work has been developed to address changes in elastic behavior when strains are not small. Such approaches must address two issues: (i) The importance of finite deformation measures beyond the infinitesimal-strain limit, i.e., descriptions of strain that go beyond a symmetrized gradient of the deformation field; (ii) Modifications in the constitutive relation of the material, both in the sense of changing the variable that stresses are dependent on from infinitesimal strain measures to finite-deformation measures, and in the sense of modifying the functional form of stress-strain dependence to include explicit nonlinearities or other features motivated by the behavior of the material at larger deformations.

Tissues and cells are very complex materials, and the uncertainties translating their properties into the language of continuum mechanics become greater when going beyond infinitesimal-strain linear elasticity. It cannot be expected that any particular nonlinear model captures all salient behavior of these biological systems. We are therefore aiming at testing whether the main conclusions drawn from the controlled, analytic linear elasticity approach hold up when translating to a typical approach in nonlinear, finite-strain elasticity. As an indicative formalism, we choose quadratic Hencky elasticity, in which the elastic energy remains quadratically dependent on the strain measure (the energy keeps the quadratic form of linear elasticity, now in the logarithmic strain), but the strain measure is logarithmic in deformation gradients, accounting for finite deformations.

With $\mathbf u(\mathbf X)$ the displacement of the material point at reference position $\mathbf X$, the deformation gradient is $\mathbf F=\mathbf I+\nabla\mathbf u$~\cite{si:Bower2009,si:Papadopoulos2020}, where $\nabla$ is the gradient with respect to $\mathbf X$.
The Hencky strain is defined as~\cite{si:Hencky1928,si:Neff2016}
\begin{equation}
 \mathbf h=\tfrac12\log\bigl(\mathbf F^{\mathsf T}\mathbf F\bigr).
 \label{si:eq:hencky-strain}
\end{equation}
Because $\mathbf F^{\mathsf T}\mathbf F=\mathbf I$ for a rigid rotation, $\mathbf h$ vanishes for rotations of any size, and for small displacement gradients it reduces to the infinitesimal strain, $\mathbf h=\boldsymbol\epsilon+O(|\nabla\mathbf u|^2)$.

The energy density per unit reference volume in region $s$ is then~\cite{si:Hencky1929,si:Anand1979}
\begin{equation}
 w_s=G_s\operatorname{tr}\bigl(\mathbf h^2\bigr)+\tfrac12\bigl(K_s-\tfrac23G_s\bigr)(\operatorname{tr}\mathbf h)^2 ,
 \label{si:eq:hencky-density}
\end{equation}
where $K_s$ and $G_s$ are the bulk and shear moduli of disk and ring.
For small strains $w_s$ becomes the energy density of linear elasticity, so the two theories share these moduli.
As in linear elasticity (Section S5), the displacement and the scale $\Lambda$ minimize the total energy $U=\sum_s\int w_s\,dV$ over the reference cell, with the outline scaled by $\Lambda$.
Here the minimization couples the Fourier modes, so we solve it by finite elements.

\begin{figure}[tbp]
  \centering
  \includegraphics[width=0.75\linewidth]{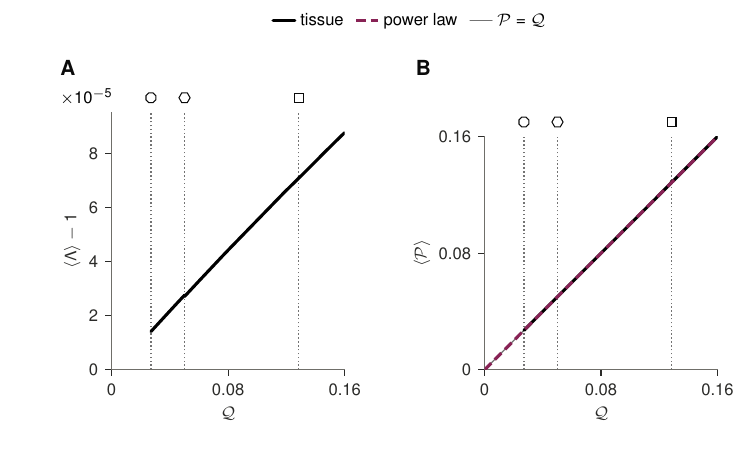} %
  \caption{(A) Average scale factor $\Lambda$  between cells of equilibrium area and cells of nominal unstrained area under quadratic Hencky elasticity. The deviations from 1 are much smaller than in linear elasticity (see Fig.~\ref{si:fig:cellareas}), and variation between cell topologies are much smaller (not shown). (B) As a consequence, the averaged perimeter excess is almost identical to the shape excess under quadratic Hencky elasticity (cf.\ Fig.~\ref{si:fig:perimetermeasures}).
  }
  \label{si:fig:henckyap}
\end{figure}

One advantage of such an approach is that the contributions from deformation Fourier modes $f_n$ with $n\ne 0$ do retain an explicit area dependence (which is entirely decoupled in the infinitesimal-strain formalism), so that the physically meaningful near-conservation of cross-sectional area for soft materials under plane strain is more strongly enforced. Figure~\ref{si:fig:henckyap}A demonstrates that, for the entire range of deformations considered, the deformed equilibrium area $A_{eq}$ deviates from the unstrained area $A_0$ by less than one percent. As a consequence, the slight difference between the measures of perimeter excess ${\cal P}$ and shape excess ${\mathcal Q}$ becomes vanishingly small (Fig.~\ref{si:fig:henckyap}B).

\begin{figure}[tbp]
  \centering
      \includegraphics[width=0.7\linewidth]{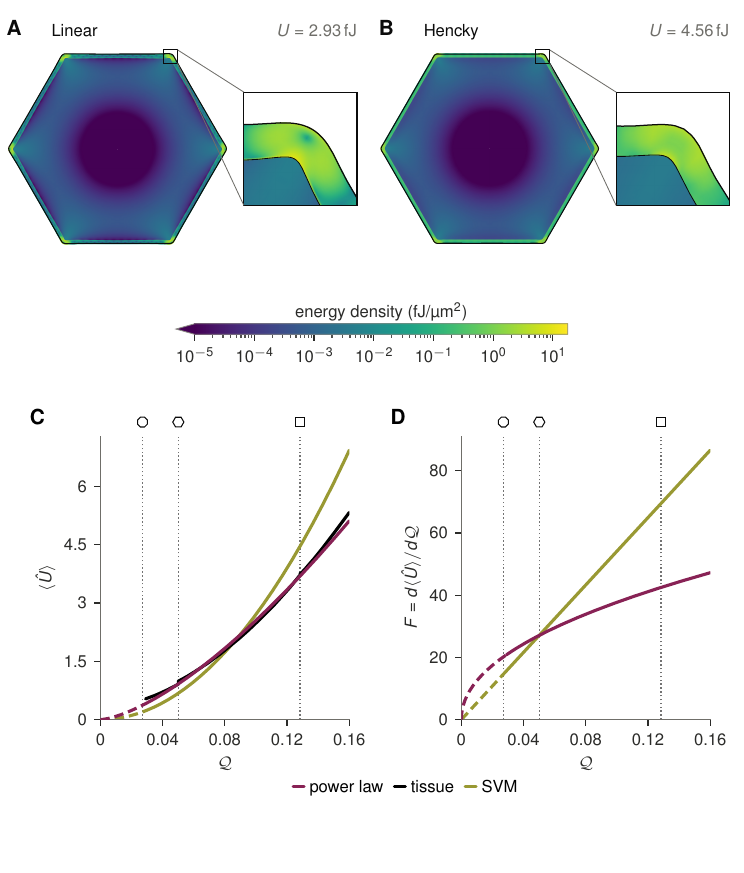} %
  \caption{(A,B) Comparison of deformation energy density maps for a regular-hexagon deformation under linear elasticity (A) and quadratic Hencky elasticity (B). Distributions are very similar, with somewhat stronger cortical concentration in (B). (C) Best-fit of power laws (SVM and general power) to $\hat{U}({\mathcal Q})$ for quadratic Hencky elasticity in analogy to Fig.~4C of the main text, again showing subquadratic behavior. (D) Derivatives of the curves in (C), preserving the conclusion of slowly varying forces in Hencky nonlinear elasticity.
  }
  \label{si:fig:henckyenergy}
\end{figure}

The approach of quadratic Hencky elasticity takes into account the nonlinear strain terms generated by rotational deformations, which vanish for small displacement gradients. Because the cortex rotates substantially and is nearly incompressible, these terms raise the total energy considerably. Rather than comparing the absolute values of energy, we point out that in both theories the energy density is concentrated in the cortex (Fig.~\ref{si:fig:henckyenergy}(A,B)) and that the dependence of total deformation energy on the shape parameter ${\mathcal Q}$ remains subquadratic, with $\alpha_2=1.48$ (Fig.~\ref{si:fig:henckyenergy}C). The salient conclusions of the main text are thus robust towards the introduction of nonlinear effects. More quantitative results could be obtained with closer modeling of constitutive relations for the cytoplasmic and cortical materials, but while such nonlinear modeling has been attempted for tissues as a whole (as an effective material), it is not presently available on the cellular or subcellular scale.

\end{document}